\documentclass[11pt]{article}
\usepackage[margin=1in]{geometry}

\usepackage{graphicx}%
\usepackage{amsmath,amssymb,amsfonts}%
\usepackage{amsthm}%
\usepackage{bm}
\usepackage[title]{appendix}%
\usepackage{enumitem}
\usepackage{todonotes}
\usepackage{bbm}
\usepackage{lscape}
\usepackage{multirow}
\usepackage{array}
\usepackage{cleveref}
\usepackage{booktabs}%
\usepackage{threeparttable}
\usepackage{helvet}
\usepackage{authblk}
\usepackage[round]{natbib}
\usepackage{setspace}

\title{Small Samples and Short Panels: Evaluating Policy Evaluation Methods with Realistic Data}

\author[1]{Luke Stewart}
\author[2]{Gary Hettinger}
\author[3]{Youjin Lee}
\author[1]{Nandita Mitra}

\affil[1]{University of Pennsylvania}
\affil[2]{New York University}
\affil[3]{Brown University}

\begin{document}

\maketitle

\abstract{Methods for estimating causal effects in longitudinal, quasi-experimental settings are widely used in economics, public health, political science, and other fields. However, studies evaluating effects of health policies often rely on limited sample sizes, both in terms of study units and time periods analyzed. Synthetic difference-in-differences (SDiD) and the augmented synthetic control method (ASCM) are recently developed methods for evaluating the effects of policy interventions. Although SDiD and ASCM generally rely on weaker assumptions than both DiD and SCM, they lack theoretical performance guarantees with respect to bias or coverage in relevant settings with small sample sizes or short panel lengths. To evaluate the performance of SDiD and ASCM in realistic, small-sample settings, we employ a calibrated simulation strategy that allows the injection of a known treatment effect into existing data in a setting of interest. Drawing on findings from these empirical investigations, we offer practical guidance for researchers and policymakers on when SDiD and ASCM are likely to yield reliable estimates and inferences under realistic scenarios.}

\doublespacing
\section{Introduction}

Researchers frequently use causal inference methods, such as difference-in-differences (DiD) and the synthetic control method (SCM), to evaluate the effects of economic \citep{goldsmith-pinkhamTrackingCredibilityRevolution2026} and public health \citep{wangAdvancesDifferenceindifferencesMethods2024} policy interventions using panel data. These studies are often conducted at the level of geographic units, such as cities, counties, or states, rather than at the individual level. However, evaluating policy interventions across geographic units can be challenging because outcome data are often unavailable for many units or are observed over only a limited number of time periods. As a result, researchers frequently must conduct analyses using relatively few geographic units and short panel lengths. For example, state and local surveys of health behaviors, such as the Behavioral Risk Factor Surveillance System (BRFSS) and its associated Selected Metropolitan/Micropolitan Area Risk Trends (SMART), include many individual participants but cover relatively few localities and, depending on the outcome of interest, relatively few time periods. Moreover, although SMART has collected annual data since 2009, many metropolitan areas meet the inclusion criteria in only a subset of those years \citep{BehavioralRiskFactor2026}. 

The econometrics literature has extensively studied valid estimation and inference for DiD designs with a small number of units; see section 5.1 of \cite{rothWhatsTrendingDifferenceindifferences2023} for a review. Additional challenges can arise when the core assumption around DiD methods, the parallel trends assumption, may only hold conditional on a set of covariates. Under this conditional parallel trends assumption, estimation requires specifying a propensity score model \citep{abadieSemiparametricDifferenceinDifferencesEstimators2005}, an outcome regression model, or both in a doubly robust framework \citep{santannaDoublyRobustDifferenceindifferences2020}. When only a small number of units are available, however, these models can be difficult to estimate reliably, particularly when many covariates are included or flexible modeling approaches are required.

When the parallel trends assumption is unlikely to hold or is only plausible after conditioning on a large set of covariates, particularly in settings with few units, researchers may instead turn to alternative methods. These include SCM \citep{abadieEconomicCostsConflict2003, abadieSyntheticControlMethods2010}, extensions to SCM \citep{ben-michaelAugmentedSyntheticControl2021a, ben-michaelSyntheticControlsStaggered2022, xuGeneralizedSyntheticControl2017, arkhangelskySyntheticDifferenceinDifferences2021}, and matrix completion methods \citep{atheyMatrixCompletionMethods2021}. Unlike DiD, however, the theoretical guarantees for estimation and inference with SCM and related methods, summarized in \cref{background}, generally rely on the availability of many pre-intervention outcome measurements. This requirement can be difficult to satisfy in evaluations of health policy interventions, where reliable outcome data are often available for only a limited number of pre-intervention periods and only a subset of geographic units. Furthermore, even when long pre-intervention histories are available, they may be contaminated by other observed or unobserved shocks that alter units' potential outcomes, thereby weakening the validity of these methods. Despite these practical challenges, relatively little work has systematically evaluated the performance of SCM and related methods for estimating policy effects and conducting inference in settings with few units or short pre-intervention panels.

To provide practical guidance for researchers evaluating policy interventions in settings with few units, short panels, and where the parallel trends assumption is unlikely to hold, we conduct a simulation study calibrated to empirical data. We compare the performance of three extensions to SCM: synthetic difference-in-differences \citep{arkhangelskySyntheticDifferenceinDifferences2021}, the augmented synthetic control method \citep{ben-michaelAugmentedSyntheticControl2021a}, and partially pooled synthetic controls \citep{ben-michaelSyntheticControlsStaggered2022}. Specifically, we evaluate the bias, empirical standard error, and confidence interval coverage of these methods for estimating the average treatment effect on the treated (ATT) under simultaneous policy adoption. Our simulations include settings with as few as six control units and five pre-treatment periods, allowing us to assess performance under severe data limitations. We also examine how performance changes as the number of treated and control units and the lengths of the pre- and post-treatment periods increase. We begin with a brief methodological review of the competing methods, emphasizing the assumptions underlying their theoretical guarantees, particularly those related to sample size and panel length. We then describe our simulation design, including its calibration to empirical datasets, and conclude by presenting the simulation results, practical recommendations for applied researchers, and the limitations of our study.

\section{Methodological background}\label{background}
\subsection{Synthetic difference-in-differences}
Synthetic difference-in-differences (SDiD) \citep{arkhangelskySyntheticDifferenceinDifferences2021} was developed to estimate causal effects in panel data when untreated outcomes may be driven by latent factors that make the standard parallel trends assumption implausible. It combines key elements from both SCM and DiD. Like SCM, it constructs a weighted average of control unit outcomes to estimate the counterfactual average outcome that the treated units would have experienced in the post-treatment period had they not received the intervention. Two sets of weights are computed: unit weights that upweight control unit outcomes with trends more similar to the trend of the average of the treated unit outcomes in the pre-treatment period, and time weights that upweight pre-treatment periods where the average control unit outcome is most similar to the average of the control unit outcomes in the post-treatment period. Unlike in SCM, the unit weights include an offset term that allows the weighted average to serve as a counterfactual for the trend in the treated units' outcomes, instead of the level of the outcomes. Using the two sets of weights, a weighted two-way fixed effects (TWFE) regression is used to estimate the treatment effect.

Under certain modeling assumptions, the coefficient from the weighted TWFE regression is an unbiased and asymptotically normal estimator of the ATT. The primary assumption is that the outcome is generated according to a latent factor model, which we discuss in \cref{simulation_setup}.  The remaining assumptions most relevant to our study are the asymptotic assumptions used to establish the theoretical properties of the SDiD estimator. Unlike in cross-sectional settings, asymptotics in this panel setting concern both the number of units and the number of periods. Assuming that all units that adopt the policy do so in the same period, let $N_{co}$, $N_{tr}$, $T_{pre}$, and $T_{post}$ denote the number of control units, the number of treated units, the number of pre-treatment periods, and the number of post-treatment periods, respectively. Then, \cite{arkhangelskySyntheticDifferenceinDifferences2021} assume that:
\begin{enumerate}
    \item The product $N_{tr}T_{post}$ goes to infinity, and both $N_{co}$ and $T_{pre}$ go to infinity. \label{assump:asymptotic_1}
    \item The ratio $T_{pre}/N_{co}$ is bounded and bounded away from zero.\label{assump:asymptotic_2}
    \item The quantity $\frac{N_{co}}{\log^2(N_{co})} \cdot \frac{1}{N_{tr}T_{post} \max (N_{tr}, T_{post})}$ goes to infinity.\label{assump:asymptotic_3}
\end{enumerate}

The first point requires that both the number of control units and pre-treatment periods goes to infinity, and that either the number of treated units or post-treatment periods goes to infinity. In practice, this assumption allows the number of treated units or the number of post-treatment periods to be small, but not both. The second point requires that the number of control units and the number of pre-treatment periods grow at the same rate. Finally, the third point requires that the number of control units grows much faster than the number of treated unit-periods. For simplicity, suppose $T_{post}=1$ (i.e., we only observe 1 period after treatment). Then the quantity in the third point above becomes
\begin{equation*}\label{eq: assump_2iii_rewrite}
    \frac{N_{co}}{\log^2(N_{co})} \cdot \frac{1}{N_{tr}^2} 
\end{equation*}
If, for example, $N_{co}$ grows quadratically with $N_{tr}$, this quantity goes to 0 (instead of growing to infinity) as $N_{tr}$ goes to infinity (as is required by the first point above). Essentially, at least one of these assumptions fails to hold if there are few treated units and few post-treatment periods, if there are far more pre-treatment periods than control units or vice versa, or if the number of control units is not much larger than the number of treated observations (the product of treated units and post-treatment periods).

It is important to emphasize that the assumptions listed here as well as the other technical assumptions in \cite{arkhangelskySyntheticDifferenceinDifferences2021} are sufficient but not necessary for the estimator to be unbiased and asymptotically normal. Unlike methods in the interactive fixed effects family (e.g., \citealp{xuGeneralizedSyntheticControl2017, baiPanelDataModels2009}), SDiD does not attempt to estimate all the parameters of the latent factor model, so it may perform favorably under different data generating mechanisms and in settings where the sample size and panel length fail to satisfy the assumptions above. 

\cite{arkhangelskySyntheticDifferenceinDifferences2021} describe three methods to estimate the asymptotic variance of the SDiD estimator (each of which is included in their accompanying software), but we focus here on the placebo variance estimation method \citep{abadieSyntheticControlMethods2010, doudchenkoBalancingRegressionDifferenceInDifferences2016} as it is the most applicable when $N_{tr}$ is small and is the only method that one can use with a single treated unit. In this method, control unit outcomes are resampled many times. In each new sample, treatment is randomly assigned to $N_{tr}$ out of $N_{co}$ resampled units. SDiD estimates are then computed for each of these samples, and the sample variance of this collection of estimates is used as an estimate of the variance of the original SDiD estimate. This method assumes that the variance of the error distribution for SDiD estimators computed using the treated units is the same as that for SDiD estimators that only utilize outcomes from the control units.

\subsection{Augmented synthetic controls}
The augmented synthetic control method (ASCM) extends the original synthetic control method by correcting for imperfect pre-treatment fit using an outcome model that relates post-treatment outcomes to pre-treatment outcomes and covariates. To estimate a counterfactual outcome in settings with a single treated unit, ASCM first computes SCM weights and estimates the SCM counterfactual. Then, ASCM uses the predicted outcomes from the outcome model to assess the amount of bias induced by poor pre-treatment fit of the SCM weights. Specifically, the method computes the SCM-weighted average of predictions for the post-treatment outcomes in the control units from the outcome model and subtracts the model prediction of the post-treatment outcome in the treated unit. This difference represents an estimate of the bias induced by a poor pre-treatment fit of the SCM weights. The ASCM estimate of the treated unit's counterfactual outcome is the SCM estimate minus the bias term computed using the outcome model. When multiple treated units are present, the average of their outcomes is computed in each period and used as a single ``pooled'' outcome series.

To test hypotheses and construct confidence intervals, \cite{ben-michaelAugmentedSyntheticControl2021a} utilize a conformal inference procedure developed in \cite{chernozhukovExactRobustConformal2021} to compute the p-value of a given null hypothesis. Residuals are computed by subtracting the ASCM fit from the observed outcome in the treated unit. When the post-treatment residuals are more extreme relative to the pre-treatment residuals (in magnitude), the p-value will be smaller. However it is important to note that the computed p-value will not be smaller than $\frac{1}{T}$ under this procedure, with $T$ being the total number of periods. Confidence intervals can be obtained through test inversion. \cite{ben-michaelAugmentedSyntheticControl2021a} show that the p-value resulting from this procedure is asymptotically valid as the number of pre-treatment periods grows.

In theory, any outcome model can be used to correct the bias resulting from the SCM method, but \cite{ben-michaelAugmentedSyntheticControl2021a} show that using a ridge regression of post-treatment outcomes on pre-treatment outcomes yields favorable statistical properties. Under a latent factor model, they derive a finite sample error bound for ASCM that holds with high probability. This bound depends on, among other things, the imbalance in the pre-treatment outcomes between treated and control units, scaled by $T_{pre}^{-1/2}$. Therefore, a long pre-treatment period will tend to reduce the error of the ASCM estimator. 

\subsection{Partially pooled synthetic controls}
Partially pooled synthetic controls \citep{ben-michaelSyntheticControlsStaggered2022} extends SCM to settings with multiple treated units, particularly those with staggered treatment adoption, by combining information across treated units when constructing synthetic controls. In contrast to ASCM, which constructs a counterfactual for the average outcome of the treated units, partially pooled SCM allows separate counterfactuals to be estimated for each treated unit. Therefore, instead of estimating $N_{co}$ different weights, partially pooled SCM estimates $N_{co}\cdot N_{tr}$ different weights. The weights are estimated by minimizing a weighted average of two objective functions: one that measures error in the pre-treatment fit separately for each treated unit, and one that measures error in the pre-treatment fit for the average of the treated units (the ``pooled'' pre-treatment fit). The pooled pre-treatment fit receives weight $\nu$, and the separate pre-treatment fit receives weight $1-\nu$, where $\nu$ is a user-chosen parameter between 0 and 1. At one extreme, $\nu=1$ corresponds to SCM applied to the average outcome of the treated units, which is equivalent to ASCM with no outcome model augmentation. On the other hand, $\nu=0$ corresponds to conducting separate SCM for each treated unit and averaging the ATT estimates together. 

In the case of staggered adoption, under a linear factor model, \cite{ben-michaelSyntheticControlsStaggered2022} derive finite sample error bounds for estimating the ATT that depend both on the quality of the pooled pre-treatment fit and the separate pre-treatment fit. However, they show that under simultaneous adoption, the finite sample error bound does not depend on the separate pre-treatment fit. Therefore, in settings with simultaneous adoption, it is theoretically favorable to choose $\nu=1$ and simply conduct SCM for the average outcome of the treated units. We confirm that this theoretical result holds through simulation and compare the performance of ASCM and partially pooled SCM in \Cref{part_pool}.

\section{Comparison of methods for studies with small samples and short panels}\label{simulation_setup}
To evaluate the performance of SDiD and ASCM in settings with limited numbers of units and time periods, we conduct a simulation study using a data-generating process calibrated to real-world data. We simulate datasets with varying sample sizes and panel lengths to isolate the effects of these design characteristics on estimator performance. Let $Y_{it}$ denote the observed outcome for unit $i$ in period $t$, and let $W_{it}$ denote a binary treatment indicator. We generate the data under the following latent factor model, for which both SDiD and ASCM have favorable theoretical properties:
\begin{equation}\label{eq:model}
    \bm{Y}=\bm{L} + \bm{W \cdot \tau} + \bm{E} = \bm{\Gamma \Psi}' + \bm{W \cdot \tau} + \bm{E}
\end{equation}
where $\bm{Y}$ is an $N \times T$ matrix with entries $Y_{it}$, $\bm{W \cdot \tau}$ is an $N \times T$ matrix with entries $W_{it} \tau_{it}$, $\tau_{it}$ is the treatment effect for unit $i$ in period $t$, and $\bm{E}$ is an $N \times T$ matrix of errors. $\bm{L}$ is a fixed matrix which can be decomposed into an $N \times r$ matrix $\bm{\Gamma}$ of factor loadings that vary across units and an $T \times r$ matrix $\bm{\Psi}$ of factors that vary over time. 

To ensure that our simulations reflect realistic settings, we estimate $\bm{\Gamma}$, $\bm{\Psi}$, and the distribution of $\bm{E}$ from two real-world datasets (referred to as the \emph{calibration datasets}). These estimates are then used to generate simulated datasets with varying sample sizes and panel lengths. We compute estimates $\hat{\bm{\Gamma}}$ and $\hat{\bm{\Psi}}$ using the method of interactive fixed effects \citep{baiPanelDataModels2009}. We then fit a multivariate Gaussian distribution to $\hat{\bm{\Gamma}}$. For a chosen sample size $N$, we first simulate $\bm{L}$ by drawing $N$ factor loadings from the Gaussian fit to $\hat{\bm{\Gamma}}$ and multiplying by $\hat{\bm{\Psi}}$. If the desired panel length $T$ is less than the length of the calibration dataset, we use the $T$ most recent periods of $\hat{\bm{\Psi}}$. For each round of the simulation, we sample the rows of $\bm{E}$ from a multivariate Gaussian, which, along with the simulated matrix $\bm{L}$, yields the simulated outcomes $\bm{Y}$. We then randomly assign treatment to $N_{tr}$ units, allowing the probability of being treated to be higher in units with systematically larger outcomes (i.e., larger entries of $\bm{L}$). For all simulations, we set the treatment effect to zero ($\tau_{it}=0$). Consequently, the simulations evaluate the bias and variability of SDiD and ASCM under the null, as well as the coverage of their corresponding confidence intervals. Additional details on the calibration procedure and data generation are provided in \Cref{additional_sim}.

The first calibration dataset is based on city-level youth soda consumption data from the Youth Risk Behavior Surveillance System (YRBSS), reflecting the type of public health surveillance data commonly used to evaluate beverage tax policies \citep{cdcYouthRiskBehavior2026}. From YRBSS microdata, we compute mean rates of soda consumption (measured in number of servings consumed per week) among high schoolers in $N=7$ cities, every 2 years from 2007 to 2017 (giving $T=6$). The second calibration dataset consists of average log wages by state and year from the Current Population Survey (CPS), first considered in \cite{bertrandHowMuchShould2004} and used again in \cite{arkhangelskySyntheticDifferenceinDifferences2021}. This dataset consists of observations from $N=50$ states across $T=40$ years. Because this dataset contains a relatively long panel, it allows us to investigate how the performance of SDiD and ASCM changes as panel length increases.

In addition to comparing SDiD and ASCM under the latent factor model, we compare both methods with traditional DiD in a setting where parallel trends holds to quantify the loss in precision associated with using methods that are robust to violations of the parallel trends assumption. We generate data using a two-way fixed effects model in which the unit and time fixed effects are estimated from the CPS calibration dataset. Specifically, the time fixed effects are taken from the first dimension of the estimated factors and the unit fixed effects are taken from the second dimension of the estimated factor loadings. To account for small sample sizes in DiD estimation, we follow the approach outlined in \cite{leeSimpleApproachesInference2025}. Specifically, we collapse the time dimension to pre-treatment and post-treatment averages of the outcomes, compute the differences between these, and regress the difference in averages on an indicator for being treated. When $N_{tr}=1$, we utilize the classical least squares standard error. When $N_{tr}>1$, we utilize robust standard errors.

For each simulated dataset, we estimate the ATT using SDiD, implemented in the \texttt{synthdid} R package, and ASCM, implemented in the \texttt{augsynth} R package with ridge regression as the outcome model.\footnote{In preliminary investigations, we found that the standard errors computed by the \texttt{synthdid} package vastly underestimated the true standard deviation of the sampling distribution of the SDiD estimator. We believe this is due to an issue with the preliminary optimization settings for the placebo control method of estimating the asymptotic variance in the version of \texttt{synthdid} that we used. We thank Petra Tschuchnig for providing code that fixes this issue.}

\section{Simulation results}
For each simulation scenario, we summarize estimator performance using bias, empirical standard error, and confidence interval coverage, together with their Monte Carlo standard errors \citep{morrisUsingSimulationStudies2019}. As such, we ran 1,900 simulations for each specification, which ensures that the standard error of the coverage rate is no more than 1 percentage point when 95\% coverage is achieved. To aid interpretability, the bias and empirical standard error of each estimator is divided by the standard deviation of the outcome after demeaning the outcome within each year. For example, a reported bias of 0.1 indicates that the average bias of the estimator is one tenth the standard deviation of the outcome across units within a given year. Coverage was calculated as 100\% minus the proportion of simulations in which the true null of $\tau=0$ was rejected.

\Cref{tab:soda} summarizes the simulation results and corresponding Monte Carlo standard errors for the scenarios calibrated to the soda consumption data. Because this dataset contains only five pre-treatment and one post-treatment periods, we consider only this panel length while varying the number of treated units from 1 to 25 and the number of control units from 6 to 40. Despite the limited sample size, both SDiD and ASCM exhibited moderate bias, with the magnitude of the bias almost always remaining below 0.1 empirical standard deviations. ASCM generally produced smaller bias than SDiD, particularly when more treated and control units were available. As expected, the empirical standard error of both estimators decreased as the numbers of treated and control units increased, with larger gains from increasing the number of treated units. Across nearly all simulation settings, ASCM also exhibited lower empirical standard error than SDiD. The two methods differed more substantially in confidence interval performance. As expected when $T<20$, the conformal confidence intervals for ASCM consistently overcovered the true treatment effect. Although ASCM exhibited lower empirical standard error than SDiD, this improved precision was not reflected in the conformal intervals. In contrast, SDiD achieved approximately nominal coverage once at least 25 control units were available with a single treated unit or at least 10 control units were available with six treated units.

\Cref{tab:cps} shows the corresponding summary estimates and standard errors for the simulations calibrated to the CPS wage data. For brevity, we display fewer combinations of $N_{co}$ and $N_{tr}$ here. In comparison to SDiD, ASCM exhibits larger bias when there are more time periods and few treated units. With fewer periods or more treated units, the biases of the two estimators are comparable. SDiD has lower standard errors than ASCM for all values of $N$ and $T$ in this set of simulations. In an extreme case with $N_{tr}=1, N_{co}=10, T_{pre}=30$, and $T_{post}=10$, the variance of ASCM is 2.26 times larger than that of SDiD. SDiD features similar coverage as in the soda consumption simulations. When only one treated unit is available, with 10 or fewer control units, the intervals from this estimator undercover slightly, with coverage ranging from 85\%--90\%, regardless of the length of the panel. With more units (more than 10 control units or at least six treated units), this estimator achieves close to nominal coverage. ASCM nearly achieves nominal coverage for any number of units, as long as at least 20 periods are available. 

\Cref{tab:cps_par} summarizes the simulation results under the parallel trends assumption, allowing us to quantify the loss in precision associated with using SDiD and ASCM when traditional DiD is correctly specified. For instance, relative efficiency (calculated as a ratio of estimator variance to DiD variance) ranges from 1.09 to 2.25 for SDiD and ranges from 1.22 to 1.84 for ASCM. For most settings, the relative efficiency of both SDiD and ASCM fall below 1.5. SDiD's performance relative to DiD improves as the number of control units increases and worsens as the total panel length increases. There is no change in its relative performance as the number of treated units increases. The relative efficiency of ASCM improves with more control units, but can worsen as the number of treated units grows. There are no consistent patterns in relative performance for ASCM as the panel length changes. Similar patterns in coverage persist from the other data generating processes. SDiD undercovers when few units are available. ASCM overcovers when few time periods are available, but achieves nominal coverage for $T\geq 20$. Unsurprisingly, DiD achieves nominal coverage regardless of sample size. 

\begin{table}
\centering
\scalebox{0.8}{
\begin{threeparttable}
\caption{Performance metrics for simulations calibrated to YRBSS soda consumption data}
\label{tab:soda}
\begin{tabular}{llcccccc}
\multicolumn{8}{c}{$\bm{T_{pre}=5, T_{post}=1}$} \\
\toprule
& & \multicolumn{2}{c}{Bias} & \multicolumn{2}{c}{Empirical SE} & \multicolumn{2}{c}{Coverage (\%)} \\
\cmidrule(lr){3-4}\cmidrule(lr){5-6}\cmidrule(lr){7-8}
$N_{tr} $& $N_{co}$ & SDiD & ASCM & SDiD & ASCM & SDiD & ASCM \\
\midrule
\multirow{4}{*}{1} & 6 & -0.012 (0.013) & -0.014 (0.010) & 0.567 (0.009) & 0.456 (0.007) & 87.579 (0.757) & 100.000 (0.000)\\
 & 10 & -0.082 (0.012) & 0.040 (0.010) & 0.523 (0.008) & 0.439 (0.007) & 87.053 (0.770) & 100.000 (0.000)\\
 & 25 & -0.054 (0.010) & 0.006 (0.008) & 0.427 (0.007) & 0.357 (0.006) & 94.526 (0.522) & 100.000 (0.000)\\
 & 40 & -0.055 (0.009) & 0.012 (0.008) & 0.404 (0.007) & 0.339 (0.006) & 94.842 (0.507) & 100.000 (0.000)\\
\cmidrule{1-8}
\multirow{3}{*}{6} & 10 & -0.096 (0.006) & 0.019 (0.005) & 0.248 (0.004) & 0.222 (0.004) & 97.211 (0.378) & 100.000 (0.000)\\
 & 25 & -0.082 (0.005) & -0.005 (0.004) & 0.205 (0.003) & 0.172 (0.003) & 92.842 (0.591) & 100.000 (0.000)\\
 & 40 & -0.071 (0.004) & 0.003 (0.003) & 0.186 (0.003) & 0.151 (0.002) & 93.526 (0.565) & 100.000 (0.000)\\
\cmidrule{1-8}
\multirow{2}{*}{10} & 25 & -0.074 (0.004) & 0.005 (0.003) & 0.160 (0.003) & 0.148 (0.002) & 94.789 (0.510) & 100.000 (0.000)\\
 & 40 & -0.082 (0.004) & 0.003 (0.003) & 0.159 (0.003) & 0.129 (0.002) & 92.368 (0.609) & 100.000 (0.000)\\
\cmidrule{1-8}
25 & 40 & -0.086 (0.003) & -0.001 (0.002) & 0.110 (0.002) & 0.090 (0.001) & 96.103 (0.444) & 100.000 (0.000)\\
\bottomrule
\end{tabular}%
\begin{tablenotes}
   \item The bias and empirical standard error of each estimator is divided by the standard deviation of the outcome variable after demeaning the outcome within each year. 
\end{tablenotes}
\end{threeparttable}
}
\end{table}

\clearpage\newpage

\begin{table}
\centering
\scalebox{0.8}{
\begin{threeparttable}
\caption{Performance metrics for simulations calibrated to CPS wage data}
\label{tab:cps}
\begin{tabular}{llcccccc}
\multicolumn{8}{c}{$\bm{T_{pre}=5, T_{post}=1}$} \\
\toprule
& & \multicolumn{2}{c}{Bias} & \multicolumn{2}{c}{Empirical SE} & \multicolumn{2}{c}{Coverage (\%)} \\
\cmidrule(lr){3-4}\cmidrule(lr){5-6}\cmidrule(lr){7-8}
$N_{tr} $& $N_{co}$ & SDiD & ASCM & SDiD & ASCM & SDiD & ASCM \\
\midrule
\multirow{4}{*}{1} & 6 & 0.004 (0.010) & -0.021 (0.012) & 0.449 (0.007) & 0.528 (0.009) & 86.737 (0.778) & 100.000 (0.000)\\
 & 10 & 0.014 (0.010) & 0.019 (0.011) & 0.432 (0.007) & 0.473 (0.008) & 89.474 (0.704) & 100.000 (0.000)\\
 & 25 & 0.000 (0.009) & -0.000 (0.010) & 0.383 (0.006) & 0.422 (0.007) & 93.947 (0.547) & 100.000 (0.000)\\
 & 40 & -0.001 (0.009) & -0.006 (0.010) & 0.385 (0.006) & 0.422 (0.007) & 93.947 (0.547) & 100.000 (0.000)\\
\cmidrule{1-8}
\multirow{2}{*}{10} & 25 & 0.003 (0.003) & 0.005 (0.004) & 0.140 (0.002) & 0.174 (0.003) & 96.263 (0.435) & 100.000 (0.000)\\
 & 40 & 0.005 (0.003) & -0.002 (0.003) & 0.132 (0.002) & 0.148 (0.002) & 94.629 (0.517) & 100.000 (0.000)\\
\bottomrule
\\\\
\multicolumn{8}{c}{$\bm{T_{pre}=15, T_{post}=5}$} \\
\toprule
& & \multicolumn{2}{c}{Bias} & \multicolumn{2}{c}{Empirical SE} & \multicolumn{2}{c}{Coverage (\%)} \\
\cmidrule(lr){3-4}\cmidrule(lr){5-6}\cmidrule(lr){7-8}
$N_{tr} $& $N_{co}$ & SDiD & ASCM & SDiD & ASCM & SDiD & ASCM \\
\midrule
\multirow{4}{*}{1} & 6 & -0.002 (0.006) & 0.033 (0.006) & 0.269 (0.004) & 0.268 (0.004) & 86.526 (0.783) & 95.053 (0.497)\\
 & 10 & -0.002 (0.006) & 0.022 (0.007) & 0.271 (0.004) & 0.305 (0.005) & 90.053 (0.687) & 93.158 (0.579)\\
 & 25 & -0.001 (0.005) & -0.005 (0.006) & 0.237 (0.004) & 0.249 (0.004) & 93.263 (0.575) & 94.579 (0.519)\\
 & 40 & -0.002 (0.005) & 0.000 (0.005) & 0.227 (0.004) & 0.222 (0.004) & 93.211 (0.577) & 94.421 (0.527)\\
\cmidrule{1-8}
\multirow{2}{*}{10} & 25 & -0.005 (0.002) & -0.004 (0.002) & 0.085 (0.001) & 0.094 (0.002) & 95.316 (0.485) & 94.789 (0.510)\\
 & 40 & -0.004 (0.002) & -0.000 (0.002) & 0.073 (0.001) & 0.086 (0.001) & 95.316 (0.485) & 93.632 (0.560)\\
\bottomrule
\\\\
\multicolumn{8}{c}{$\bm{T_{pre}=30, T_{post}=10}$} \\
\toprule
& & \multicolumn{2}{c}{Bias} & \multicolumn{2}{c}{Empirical SE} & \multicolumn{2}{c}{Coverage (\%)} \\
\cmidrule(lr){3-4}\cmidrule(lr){5-6}\cmidrule(lr){7-8}
$N_{tr} $& $N_{co}$ & SDiD & ASCM & SDiD & ASCM & SDiD & ASCM \\
\midrule
\multirow{4}{*}{1} & 6 & -0.027 (0.009) & 0.071 (0.010) & 0.376 (0.006) & 0.428 (0.007) & 89.421 (0.706) & 91.053 (0.655)\\
 & 10 & -0.024 (0.007) & -0.066 (0.010) & 0.300 (0.005) & 0.455 (0.007) & 90.579 (0.670) & 95.895 (0.455)\\
 & 25 & -0.007 (0.005) & 0.027 (0.007) & 0.237 (0.004) & 0.312 (0.005) & 93.632 (0.560) & 92.737 (0.595)\\
 & 40 & 0.000 (0.005) & -0.021 (0.006) & 0.218 (0.004) & 0.260 (0.004) & 94.421 (0.527) & 94.158 (0.538)\\
\cmidrule{1-8}
\multirow{2}{*}{10} & 25 & 0.001 (0.002) & 0.004 (0.002) & 0.087 (0.001) & 0.098 (0.002) & 97.211 (0.378) & 94.526 (0.522)\\
 & 40 & -0.001 (0.002) & 0.000 (0.002) & 0.078 (0.001) & 0.090 (0.001) & 95.737 (0.463) & 94.316 (0.531)\\
\bottomrule
\end{tabular}
\begin{tablenotes}
   \item The bias and empirical standard error of each estimator is divided by the standard deviation of the outcome variable after demeaning the outcome within each year. 
\end{tablenotes}
\end{threeparttable}
}
\end{table}

\begin{landscape}
\begin{table}
\centering
\scalebox{0.8}{
\begin{threeparttable}
\caption{Performance metrics for simulations calibrated to CPS wage data with parallel trends}
\label{tab:cps_par}
\begin{tabular}{llcccccccccc}
\multicolumn{11}{c}{$\bm{T_{pre}=5, T_{post}=1}$} \\
\toprule
& & \multicolumn{3}{c}{Bias} & \multicolumn{3}{c}{Empirical SE} & \multicolumn{3}{c}{Coverage (\%)} \\
\cmidrule(lr){3-5}\cmidrule(lr){6-8}\cmidrule(lr){9-11}
$N_{tr} $& $N_{co}$ & SDiD & ASCM & DiD & SDiD & ASCM & DiD & SDiD & ASCM & DiD \\
\midrule
\multirow{4}{*}{1} & 6 & -0.007 (0.010) & 0.021 (0.011) & -0.007 (0.008) & 0.433 (0.007) & 0.499 (0.008) & 0.369 (0.006) & 86.474 (0.785) & 100.000 (0.000) & 95.316 (0.485)\\
 & 10 & -0.012 (0.010) & -0.024 (0.010) & -0.009 (0.009) & 0.420 (0.007) & 0.446 (0.007) & 0.374 (0.006) & 89.895 (0.691) & 100.000 (0.000) & 94.368 (0.529)\\
 & 25 & -0.003 (0.009) & 0.011 (0.009) & -0.005 (0.008) & 0.380 (0.006) & 0.402 (0.007) & 0.352 (0.006) & 93.316 (0.573) & 100.000 (0.000) & 94.684 (0.515)\\
 & 40 & 0.012 (0.008) & 0.003 (0.009) & 0.011 (0.008) & 0.368 (0.006) & 0.392 (0.006) & 0.350 (0.006) & 94.211 (0.536) & 100.000 (0.000) & 95.211 (0.490)\\
\cmidrule{1-11}
\multirow{2}{*}{10} & 25 & -0.001 (0.003) & -0.001 (0.004) & -0.001 (0.003) & 0.135 (0.002) & 0.165 (0.003) & 0.128 (0.002) & 95.947 (0.452) & 100.000 (0.000) & 95.421 (0.480)\\
 & 40 & 0.003 (0.003) & 0.001 (0.003) & 0.002 (0.003) & 0.129 (0.002) & 0.145 (0.002) & 0.125 (0.002) & 94.474 (0.524) & 100.000 (0.000) & 94.421 (0.527)\\
\bottomrule
\\\\
\multicolumn{11}{c}{$\bm{T_{pre}=15, T_{post}=5}$} \\
\toprule
& & \multicolumn{3}{c}{Bias} & \multicolumn{3}{c}{Empirical SE} & \multicolumn{3}{c}{Coverage (\%)} \\
\cmidrule(lr){3-5}\cmidrule(lr){6-8}\cmidrule(lr){9-11}
$N_{tr} $& $N_{co}$ & SDiD & ASCM & DiD & SDiD & ASCM & DiD & SDiD & ASCM & DiD \\
\midrule
\multirow{4}{*}{1} & 6 & -0.005 (0.005) & 0.001 (0.005) & -0.003 (0.004) & 0.234 (0.004) & 0.219 (0.004) & 0.175 (0.003) & 85.526 (0.807) & 95.211 (0.490) & 94.474 (0.524)\\
 & 10 & -0.008 (0.005) & -0.013 (0.005) & -0.008 (0.004) & 0.214 (0.003) & 0.197 (0.003) & 0.163 (0.003) & 88.947 (0.719) & 94.737 (0.512) & 95.737 (0.463)\\
 & 25 & 0.003 (0.004) & -0.002 (0.004) & -0.000 (0.004) & 0.187 (0.003) & 0.182 (0.003) & 0.159 (0.003) & 93.579 (0.562) & 95.000 (0.500) & 95.526 (0.474)\\
 & 40 & 0.002 (0.004) & 0.003 (0.004) & 0.002 (0.004) & 0.182 (0.003) & 0.181 (0.003) & 0.162 (0.003) & 93.684 (0.558) & 93.579 (0.562) & 94.737 (0.512)\\
\cmidrule{1-11}
\multirow{2}{*}{10} & 25 & -0.002 (0.002) & -0.001 (0.002) & -0.001 (0.001) & 0.069 (0.001) & 0.079 (0.001) & 0.058 (0.001) & 97.000 (0.391) & 94.895 (0.505) & 95.474 (0.477)\\
 & 40 & 0.002 (0.001) & 0.004 (0.002) & 0.001 (0.001) & 0.063 (0.001) & 0.076 (0.001) & 0.057 (0.001) & 95.263 (0.487) & 95.526 (0.474) & 93.684 (0.558)\\
\bottomrule
\\\\
\multicolumn{11}{c}{$\bm{T_{pre}=30, T_{post}=10}$} \\
\toprule
& & \multicolumn{3}{c}{Bias} & \multicolumn{3}{c}{Empirical SE} & \multicolumn{3}{c}{Coverage (\%)} \\
\cmidrule(lr){3-5}\cmidrule(lr){6-8}\cmidrule(lr){9-11}
$N_{tr} $& $N_{co}$ & SDiD & ASCM & DiD & SDiD & ASCM & DiD & SDiD & ASCM & DiD \\
\midrule
\multirow{4}{*}{1} & 6 & -0.001 (0.004) & 0.000 (0.004) & -0.002 (0.003) & 0.182 (0.003) & 0.155 (0.003) & 0.118 (0.002) & 88.526 (0.731) & 95.579 (0.472) & 95.684 (0.466)\\
 & 10 & 0.002 (0.004) & -0.000 (0.003) & -0.002 (0.003) & 0.170 (0.003) & 0.146 (0.002) & 0.120 (0.002) & 86.842 (0.775) & 94.474 (0.524) & 94.263 (0.533)\\
 & 25 & 0.001 (0.003) & 0.007 (0.003) & 0.003 (0.003) & 0.143 (0.002) & 0.129 (0.002) & 0.113 (0.002) & 92.211 (0.615) & 95.684 (0.466) & 94.789 (0.510)\\
 & 40 & -0.001 (0.003) & -0.001 (0.003) & -0.002 (0.003) & 0.136 (0.002) & 0.129 (0.002) & 0.114 (0.002) & 93.632 (0.560) & 94.000 (0.545) & 94.579 (0.519)\\
\cmidrule{1-11}
\multirow{2}{*}{10} & 25 & 0.001 (0.001) & 0.001 (0.001) & 0.001 (0.001) & 0.052 (0.001) & 0.051 (0.001) & 0.042 (0.001) & 96.737 (0.408) & 93.526 (0.565) & 95.421 (0.480)\\
 & 40 & -0.000 (0.001) & 0.001 (0.001) & -0.000 (0.001) & 0.048 (0.001) & 0.051 (0.001) & 0.039 (0.001) & 95.684 (0.466) & 95.158 (0.492) & 95.526 (0.474)\\
\bottomrule
\end{tabular}
\begin{tablenotes}
   \item The bias and empirical standard error of each estimator is divided by the standard deviation of the outcome variable after demeaning the outcome within each year. 
\end{tablenotes}
\end{threeparttable}
}
\end{table}
\end{landscape}

\section{Discussion}
Researchers evaluating health policy interventions frequently use panel data from a limited number of states, counties, or municipalities. These data often feature two practical challenges: limited numbers of geographic units and short pre-intervention panels. These settings are common when analyses are conducted at the city or county level and motivate the use of methods that remain valid when the parallel trends assumption is implausible. In this article, we conducted a calibrated simulation study to evaluate the performance of SDiD and ASCM under these challenging but practically important conditions. Our findings provide guidance for researchers considering these methods either as a primary analysis or as a sensitivity analysis following a conventional DiD analysis.

Researchers may choose SDiD or ASCM as their primary analysis when subject-matter knowledge suggests that the parallel trends assumption is unlikely to hold. For example, if a policy is implemented in response to a temporary spike in the outcome during the pre-intervention period, parallel trends is unlikely to be satisfied, motivating the use of methods that relax this assumption. Alternatively, researchers conducting a primary DiD analysis may use SDiD or ASCM as sensitivity analyses to assess the robustness of their findings to potential violations of parallel trends. Other approaches, including partial identification methods \citep{rambachanMoreCredibleApproach2023} and non-inferiority tests \citep{bilinskiNothingSeeHere2026}, can similarly inform the need for sensitivity analyses based on SDiD or ASCM.

The clearest practical guidance from our simulations concerns confidence interval performance. For settings with fewer than 20 time periods, the conformal inference procedure for ASCM consistently overcovers the true treatment effect, substantially reducing power. However, with 20 or more periods, ASCM achieves approximately nominal coverage regardless of the numbers of treated and control units. For SDiD, confidence interval performance depends primarily on the number of control units. With a single treated unit, nominal coverage is attained only when at least 25 control units are available, whereas with multiple treated units, nominal coverage can be achieved with as few as 10 control units.

The performance of the point estimators is less amenable to broad recommendations because it depends on the underlying data-generating process. Across both calibration datasets, however, the normalized bias of SDiD and ASCM remains small over a wide range of sample sizes and panel lengths. Although neither estimator consistently outperforms the other with respect to bias or empirical standard error, both perform reasonably well even in settings where the asymptotic assumptions underlying their theoretical properties are unlikely to hold.

When the parallel trends assumption holds exactly, both SDiD and ASCM sacrifice precision relative to DiD, consistent with the findings of \cite{arkhangelskySyntheticDifferenceinDifferences2021}. This efficiency loss is the price paid for robustness to violations of the parallel trends assumption. Consequently, researchers should not use SDiD or ASCM as their primary analysis when the parallel trends assumption is considered plausible. However, when concerns about parallel trends exist, these methods provide valuable complementary analyses, provided their finite-sample inferential properties are taken into account.

Our simulation study has several limitations. First, we calibrated the simulations using only two real-world datasets. The differences observed between these calibration datasets suggest that a broader collection of applications would yield more general conclusions. Second, all simulations were generated under a latent factor model with simultaneous policy adoption, a setting in which both SDiD and ASCM possess favorable theoretical properties. Although a more comprehensive evaluation would consider alternative data-generating mechanisms and treatment adoption patterns, our design isolates the effects of sample size and panel length, which were the primary focus of this study. Readers interested in performance under different assumptions should consult the original methodological papers.

In addition, our implementation of SDiD and ASCM should not be interpreted as prescriptive guidance for applied analyses. We did not incorporate covariates, investigate alternative outcome models or regularization parameters for ASCM, or examine alternative variance estimation procedures for SDiD. Instead, we implemented each method using the default settings available in the corresponding software packages to reflect how these methods are most likely to be applied in practice. Researchers with the time and expertise to tailor implementation choices to a specific application may achieve different performance.

Despite these limitations, our findings provide practical guidance for applied researchers choosing between DiD, SDiD, and ASCM when only a limited number of units and time periods are available. More broadly, they highlight the need for inferential procedures that maintain reliable finite-sample performance when the asymptotic assumptions underlying existing methods are unlikely to hold. Continued methodological work on interval estimation and hypothesis testing for synthetic control methods will further improve their applicability in the small-sample settings commonly encountered in health policy evaluation.

\singlespacing
\bibliography{refs}

\appendix
\counterwithin{table}{section}

\section{Additional simulation setup details}\label{additional_sim}
Let $\bm{Y}_{sim}, \bm{L}_{sim}, \bm{W}_{sim}$, and $\bm{E}_{sim}$ denote simulated quantities. For a given $\bm{L}_{sim}$, pre-specified $\bm{\tau}$, and sample sizes, we generate data for each round of simulation by assigning treatment to $N_{tr}$ units, sampling errors for each unit (the rows of $\bm{E}_{sim}$) as Gaussian vectors, and summing the components of the model to generate $\bm{Y}_{sim}$. We calibrate the factor model to real data by first estimating $\bm{\Gamma}$ and $\bm{\Psi}$ in a real dataset by the method of interactive fixed effects (IFE) after pre-specifying the factor dimension $r$ \citep{baiPanelDataModels2009}. Essentially, this method estimates $\bm{\Gamma}$ and $\bm{\Psi}$ by minimizing a squared error loss subject to orthogonality constraints on the resulting estimates. For a given number of units $N$, we generate $\bm{L}_{sim}$ according to
\begin{gather*}
    \bm{L}_{sim}=\tilde{\bm{\Gamma}} \hat{\bm{\Psi}}' \\
    \tilde{\bm{\Gamma}} \sim MVN(\hat{\mu}_{\Gamma}, \hat{\Sigma}_{\Gamma})
\end{gather*}
where $\hat{\mu}_{\Gamma}$ and $\hat{\Sigma}_{\Gamma}$ are the sample mean vector and sample covariance matrix of the IFE estimate of $\bm{\Gamma}$ and $\hat{\bm{\Psi}}$ is the IFE estimate of $\bm{\Psi}$. While $\tilde{\bm{\Gamma}}$ is randomly generated for each simulation specification, this matrix is held constant across rounds of the simulation for a given specification. By generating $\tilde{\bm{\Gamma}}$ from a parametric distribution and using the estimate of the factors $\hat{\bm{\Psi}}$, we can vary the number of units while retaining the key time series structure of the original dataset. Errors are generated according to
\begin{equation*}
    \bm{E}_{sim,i} \sim MVN(0, \hat{\Sigma}_{e})
\end{equation*}
where $\hat{\Sigma}_e$ is an autoregressive covariance matrix fitted to the residual errors $\bm{Y}-\bm{L}_{sim}$.

To realistically assign treatment to $N_{tr}$ out of $N$ units, we compute sampling weights $w_i$ that relate to the row means of $\bm{L}_{sim}$ to reflect the fact that treatment is often associated with underlying levels of the outcome in pre-treatment periods. Specifically, we let
\begin{equation*}
    w_i = \frac{\exp(\bar{L}_i - \bar{L})}{1+\exp(\bar{L}_i - \bar{L})}
\end{equation*}
where $\bar{L}_i$ denotes the mean of the $i$th row of $\bm{L}_{sim}$ and $\bar{L}$ denotes the overall mean of $\bm{L}_{sim}$. Therefore, units with larger average values of $\bm{L}$ relative to the mean have a higher probability of receiving treatment. Units are sampled sequentially with probabilities proportional to $w_i$ until $N_{tr}$ units have been sampled, with each sampled unit being assigned as treated for the round of the simulation, yielding $\bm{W}_{sim}$.

Due to the 2017 enactment of the Philadelphia beverage tax, a tax on artificially sweetened beverages in the city of Philadelphia, fitting $\bm{L}$ to this data may result in a factor structure that incorporates a nonzero treatment effect, which may result in SDiD or ASCM estimates that are biased relative to the null treatment effect we specify in each simulation. Therefore, we compute estimates of $\bm{\Gamma}$ and $\bm{\Psi}$ from different subsets of the observed data and combine them to yield an estimate of $\bm{L}$ that is not a priori contaminated by a nonzero treatment effect. Specifically, we estimate the factor loadings $\bm{\Gamma}$ from a subset that excludes 2017, the year in which the tax went into effect, and we estimate the factors $\bm{\Psi}$ from a subset that excludes Philadelphia entirely. This process yields a factor structure that captures the important features of the data but completely rules out contamination from a nonzero true treatment effect. For both estimates, we specify a factor dimension of 2 due to the small size of the data.  For this dataset, we estimate the covariance matrix of $\bm{E}_i$ using the residuals from the IFE estimation of the factor loadings, assuming an AR(1) process.

In the CPS wage data, there are no particular treatments of interest like the Philadelphia beverage tax, so we can use estimates of the factors and factor loadings from the whole dataset to compute $\hat{\bm{\Psi}}$, $\hat{\mu}_{\Gamma}$, $\hat{\Sigma}_{\Gamma}$ and $\hat{\Sigma}_{e}$. Because this dataset is larger (in both the number of units and periods), and in line with \cite{arkhangelskySyntheticDifferenceinDifferences2021}, we use a factor dimension of 4 and an AR(2) process for the errors.

This simulation is similar to that conducted in \cite{ben-michaelAugmentedSyntheticControl2021a}, with two important differences. First, we explicitly allow for serially correlated errors across periods within units. Second, estimating $\bm{\Gamma}$ and $\bm{\Psi}$ from different subsets of the observed data allows us to calibrate our simulation to data that may contain a nonzero treatment effect, as in the soda consumption data.

\section{Results for partially pooled synthetic controls}\label{part_pool}

\Crefrange{tab:soda_part_pool}{tab:cps_par_part_pool} contain bias, empirical standard error, and coverage results for ASCM and partially pooled SCM that are analogous to \crefrange{tab:soda}{tab:cps_par} in the main text, but which exclude SDiD and DiD. The hyperparameter that balances between separate and pooled pre-treatment fits in the partially pooled SCM estimator takes two values in our table. Setting $\nu=1$ corresponds to fully pooled SCM, which is equivalent to standard SCM applied to the average outcome of the treated units. \cite{ben-michaelSyntheticControlsStaggered2022} recommend choosing $\nu$ based on the observed pre-treatment fits for a range of possible $\nu$ values, but they also provide a heuristic $\hat{\nu}$ based on the quality of the pooled fit relative to the separate fits, which we use as another potential value of $\nu$ in our simulations. This heuristic is used as the default value when using partially pooled SCM through the \texttt{augsynth} package. See section 4.2 of \cite{ben-michaelSyntheticControlsStaggered2022} for details.

\begin{landscape}
\begin{table}
\centering
\scalebox{0.75}{
\begin{threeparttable}
\caption{Performance metrics for simulations calibrated to YRBSS soda consumption data}
\label{tab:soda_part_pool}
\begin{tabular}{llccccccccc}
\multicolumn{11}{c}{$\bm{T_{pre}=5, T_{post}=1}$} \\
\toprule
& & \multicolumn{3}{c}{Bias} & \multicolumn{3}{c}{Empirical SE} & \multicolumn{3}{c}{Coverage (\%)} \\
\cmidrule(lr){3-5}\cmidrule(lr){6-8}\cmidrule(lr){9-11}
$N_{tr}$ & $N_{co}$ & ASCM & Part. pooled SCM & Part. pooled SCM & ASCM & Part. pooled SCM & Part. pooled SCM & ASCM & Part. pooled SCM & Part. pooled SCM \\
& & & $(\nu=\hat{\nu})$ & $(\nu=1)$ & & $(\nu=\hat{\nu})$ & $(\nu=1)$ & & $(\nu=\hat{\nu})$ & $(\nu=1)$ \\
\midrule
\multirow{4}{*}{1} & 6 & -0.014 (0.010) & -0.076 (0.012) & -0.076 (0.012) & 0.456 (0.007) & 0.515 (0.008) & 0.515 (0.008) & 100.000 (0.000) & 100.000 (0.000) & 100.000 (0.000)\\
 & 10 & 0.040 (0.010) & -0.050 (0.012) & -0.050 (0.012) & 0.439 (0.007) & 0.535 (0.009) & 0.535 (0.009) & 100.000 (0.000) & 100.000 (0.000) & 100.000 (0.000)\\
 & 25 & 0.006 (0.008) & -0.063 (0.011) & -0.063 (0.011) & 0.357 (0.006) & 0.473 (0.008) & 0.473 (0.008) & 100.000 (0.000) & 100.000 (0.000) & 100.000 (0.000)\\
 & 40 & 0.012 (0.008) & -0.083 (0.010) & -0.083 (0.010) & 0.339 (0.006) & 0.443 (0.007) & 0.443 (0.007) & 100.000 (0.000) & 100.000 (0.000) & 100.000 (0.000)\\
\cmidrule{1-11}
\multirow{3}{*}{6} & 10 & 0.019 (0.005) & -0.096 (0.007) & -0.105 (0.006) & 0.222 (0.004) & 0.296 (0.005) & 0.278 (0.005) & 100.000 (0.000) & 98.737 (0.256) & 98.947 (0.234)\\
 & 25 & -0.005 (0.004) & -0.086 (0.005) & -0.097 (0.005) & 0.172 (0.003) & 0.235 (0.004) & 0.235 (0.004) & 100.000 (0.000) & 100.000 (0.000) & 100.000 (0.000)\\
 & 40 & 0.003 (0.003) & -0.066 (0.005) & -0.063 (0.005) & 0.151 (0.002) & 0.204 (0.003) & 0.213 (0.003) & 100.000 (0.000) & 100.000 (0.000) & 100.000 (0.000)\\
\cmidrule{1-11}
\multirow{2}{*}{10} & 25 & 0.005 (0.003) & -0.066 (0.004) & -0.073 (0.004) & 0.148 (0.002) & 0.185 (0.003) & 0.189 (0.003) & 100.000 (0.000) & 99.947 (0.053) & 99.895 (0.074)\\
 & 40 & 0.003 (0.003) & -0.072 (0.004) & -0.073 (0.004) & 0.129 (0.002) & 0.176 (0.003) & 0.189 (0.003) & 100.000 (0.000) & 100.000 (0.000) & 100.000 (0.000)\\
\cmidrule{1-11}
25 & 40 & -0.001 (0.002) & -0.077 (0.003) & -0.076 (0.003) & 0.090 (0.001) & 0.119 (0.002) & 0.136 (0.002) & 100.000 (0.000) & 95.471 (0.477) & 95.682 (0.466)\\
\bottomrule
\end{tabular}%
\begin{tablenotes}
   \item The bias and empirical standard error of each estimator is divided by the standard deviation of the outcome variable after demeaning the outcome within each year. 
\end{tablenotes}
\end{threeparttable}
}
\end{table}
\end{landscape}

\clearpage\newpage

\begin{landscape}
\begin{table}
\centering
\scalebox{0.75}{
\begin{threeparttable}
\caption{Performance metrics for simulations calibrated to CPS wage data}
\label{tab:cps_part_pool}
\begin{tabular}{llccccccccc}
\multicolumn{11}{c}{$\bm{T_{pre}=5, T_{post}=1}$} \\
\toprule
& & \multicolumn{3}{c}{Bias} & \multicolumn{3}{c}{Empirical SE} & \multicolumn{3}{c}{Coverage (\%)} \\
\cmidrule(lr){3-5}\cmidrule(lr){6-8}\cmidrule(lr){9-11}
$N_{tr}$ & $N_{co}$ & ASCM & Part. pooled SCM & Part. pooled SCM & ASCM & Part. pooled SCM & Part. pooled SCM & ASCM & Part. pooled SCM & Part. pooled SCM \\
& & & $(\nu=\hat{\nu})$ & $(\nu=1)$ & & $(\nu=\hat{\nu})$ & $(\nu=1)$ & & $(\nu=\hat{\nu})$ & $(\nu=1)$ \\
\midrule
\multirow{4}{*}{1} & 6 & -0.021 (0.012) & 0.009 (0.010) & 0.009 (0.010) & 0.528 (0.009) & 0.429 (0.007) & 0.429 (0.007) & 100.000 (0.000) & 100.000 (0.000) & 100.000 (0.000)\\
 & 10 & 0.019 (0.011) & 0.009 (0.010) & 0.009 (0.010) & 0.473 (0.008) & 0.441 (0.007) & 0.441 (0.007) & 100.000 (0.000) & 100.000 (0.000) & 100.000 (0.000)\\
 & 25 & -0.000 (0.010) & 0.004 (0.010) & 0.004 (0.010) & 0.422 (0.007) & 0.415 (0.007) & 0.415 (0.007) & 100.000 (0.000) & 100.000 (0.000) & 100.000 (0.000)\\
 & 40 & -0.006 (0.010) & -0.006 (0.009) & -0.006 (0.009) & 0.422 (0.007) & 0.412 (0.007) & 0.412 (0.007) & 100.000 (0.000) & 100.000 (0.000) & 100.000 (0.000)\\
\cmidrule{1-11}
\multirow{2}{*}{10} & 25 & 0.005 (0.004) & 0.001 (0.004) & 0.001 (0.004) & 0.174 (0.003) & 0.157 (0.003) & 0.163 (0.003) & 100.000 (0.000) & 100.000 (0.000) & 99.632 (0.139)\\
 & 40 & -0.002 (0.003) & 0.005 (0.003) & 0.005 (0.003) & 0.148 (0.002) & 0.140 (0.002) & 0.151 (0.002) & 100.000 (0.000) & 100.000 (0.000) & 100.000 (0.000)\\
\bottomrule
\\\\
\multicolumn{11}{c}{$\bm{T_{pre}=15, T_{post}=5}$} \\
\toprule
& & \multicolumn{3}{c}{Bias} & \multicolumn{3}{c}{Empirical SE} & \multicolumn{3}{c}{Coverage (\%)} \\
\cmidrule(lr){3-5}\cmidrule(lr){6-8}\cmidrule(lr){9-11}
$N_{tr}$ & $N_{co}$ & ASCM & Part. pooled SCM & Part. pooled SCM & ASCM & Part. pooled SCM & Part. pooled SCM & ASCM & Part. pooled SCM & Part. pooled SCM \\
& & & $(\nu=\hat{\nu})$ & $(\nu=1)$ & & $(\nu=\hat{\nu})$ & $(\nu=1)$ & & $(\nu=\hat{\nu})$ & $(\nu=1)$ \\
\midrule
\multirow{4}{*}{1} & 6 & 0.033 (0.006) & 0.000 (0.005) & 0.000 (0.005) & 0.268 (0.004) & 0.216 (0.003) & 0.216 (0.003) & 95.053 (0.497) & 100.000 (0.000) & 100.000 (0.000)\\
 & 10 & 0.022 (0.007) & 0.007 (0.006) & 0.007 (0.006) & 0.305 (0.005) & 0.255 (0.004) & 0.255 (0.004) & 93.158 (0.579) & 100.000 (0.000) & 100.000 (0.000)\\
 & 25 & -0.005 (0.006) & -0.001 (0.005) & -0.001 (0.005) & 0.249 (0.004) & 0.235 (0.004) & 0.235 (0.004) & 94.579 (0.519) & 100.000 (0.000) & 100.000 (0.000)\\
 & 40 & 0.000 (0.005) & -0.004 (0.005) & -0.004 (0.005) & 0.222 (0.004) & 0.233 (0.004) & 0.233 (0.004) & 94.421 (0.527) & 100.000 (0.000) & 100.000 (0.000)\\
\cmidrule{1-11}
\multirow{2}{*}{10} & 25 & -0.004 (0.002) & -0.006 (0.002) & -0.006 (0.002) & 0.094 (0.002) & 0.096 (0.002) & 0.088 (0.001) & 94.789 (0.510) & 99.895 (0.074) & 99.947 (0.053)\\
 & 40 & -0.000 (0.002) & -0.004 (0.002) & -0.004 (0.002) & 0.086 (0.001) & 0.088 (0.001) & 0.078 (0.001) & 93.632 (0.560) & 100.000 (0.000) & 100.000 (0.000)\\
\bottomrule
\\\\
\multicolumn{11}{c}{$\bm{T_{pre}=30, T_{post}=10}$} \\
\toprule
& & \multicolumn{3}{c}{Bias} & \multicolumn{3}{c}{Empirical SE} & \multicolumn{3}{c}{Coverage (\%)} \\
\cmidrule(lr){3-5}\cmidrule(lr){6-8}\cmidrule(lr){9-11}
$N_{tr}$ & $N_{co}$ & ASCM & Part. pooled SCM & Part. pooled SCM & ASCM & Part. pooled SCM & Part. pooled SCM & ASCM & Part. pooled SCM & Part. pooled SCM \\
& & & $(\nu=\hat{\nu})$ & $(\nu=1)$ & & $(\nu=\hat{\nu})$ & $(\nu=1)$ & & $(\nu=\hat{\nu})$ & $(\nu=1)$ \\
\midrule
\multirow{4}{*}{1} & 6 & 0.071 (0.010) & 0.014 (0.010) & 0.014 (0.010) & 0.428 (0.007) & 0.431 (0.007) & 0.431 (0.007) & 91.053 (0.655) & 100.000 (0.000) & 100.000 (0.000)\\
 & 10 & -0.066 (0.010) & -0.021 (0.008) & -0.021 (0.008) & 0.455 (0.007) & 0.336 (0.005) & 0.336 (0.005) & 95.895 (0.455) & 100.000 (0.000) & 100.000 (0.000)\\
 & 25 & 0.027 (0.007) & -0.019 (0.007) & -0.019 (0.007) & 0.312 (0.005) & 0.307 (0.005) & 0.307 (0.005) & 92.737 (0.595) & 100.000 (0.000) & 100.000 (0.000)\\
 & 40 & -0.021 (0.006) & -0.004 (0.006) & -0.004 (0.006) & 0.260 (0.004) & 0.279 (0.005) & 0.279 (0.005) & 94.158 (0.538) & 100.000 (0.000) & 100.000 (0.000)\\
\cmidrule{1-11}
\multirow{2}{*}{10} & 25 & 0.004 (0.002) & 0.002 (0.002) & -0.001 (0.002) & 0.098 (0.002) & 0.097 (0.002) & 0.098 (0.002) & 94.526 (0.522) & 100.000 (0.000) & 100.000 (0.000)\\
 & 40 & 0.000 (0.002) & -0.003 (0.002) & -0.007 (0.002) & 0.090 (0.001) & 0.088 (0.001) & 0.084 (0.001) & 94.316 (0.531) & 100.000 (0.000) & 100.000 (0.000)\\
\bottomrule
\end{tabular}
\begin{tablenotes}
   \item The bias and empirical standard error of each estimator is divided by the standard deviation of the outcome variable after demeaning the outcome within each year. 
\end{tablenotes}
\end{threeparttable}
}
\end{table}
\end{landscape}

\begin{landscape}
\begin{table}
\centering
\scalebox{0.75}{
\begin{threeparttable}
\caption{Performance metrics for simulations calibrated to CPS wage data with parallel trends}
\label{tab:cps_par_part_pool}
\begin{tabular}{llccccccccc}
\multicolumn{11}{c}{$\bm{T_{pre}=5, T_{post}=1}$} \\
\toprule
& & \multicolumn{3}{c}{Bias} & \multicolumn{3}{c}{Empirical SE} & \multicolumn{3}{c}{Coverage (\%)} \\
\cmidrule(lr){3-5}\cmidrule(lr){6-8}\cmidrule(lr){9-11}
$N_{tr}$ & $N_{co}$ & ASCM & Part. pooled SCM & Part. pooled SCM & ASCM & Part. pooled SCM & Part. pooled SCM & ASCM & Part. pooled SCM & Part. pooled SCM \\
& & & $(\nu=\hat{\nu})$ & $(\nu=1)$ & & $(\nu=\hat{\nu})$ & $(\nu=1)$ & & $(\nu=\hat{\nu})$ & $(\nu=1)$ \\
\midrule
\multirow{4}{*}{1} & 6 & 0.021 (0.011) & -0.001 (0.010) & -0.001 (0.010) & 0.499 (0.008) & 0.433 (0.007) & 0.433 (0.007) & 100.000 (0.000) & 100.000 (0.000) & 100.000 (0.000)\\
 & 10 & -0.024 (0.010) & -0.015 (0.010) & -0.016 (0.010) & 0.446 (0.007) & 0.425 (0.007) & 0.425 (0.007) & 100.000 (0.000) & 100.000 (0.000) & 100.000 (0.000)\\
 & 25 & 0.011 (0.009) & -0.009 (0.009) & -0.009 (0.009) & 0.402 (0.007) & 0.406 (0.007) & 0.406 (0.007) & 100.000 (0.000) & 100.000 (0.000) & 100.000 (0.000)\\
 & 40 & 0.003 (0.009) & 0.006 (0.009) & 0.006 (0.009) & 0.392 (0.006) & 0.401 (0.007) & 0.401 (0.007) & 100.000 (0.000) & 100.000 (0.000) & 100.000 (0.000)\\
\cmidrule{1-11}
\multirow{2}{*}{10} & 25 & -0.001 (0.004) & -0.002 (0.004) & -0.002 (0.004) & 0.165 (0.003) & 0.154 (0.002) & 0.162 (0.003) & 100.000 (0.000) & 99.895 (0.074) & 99.895 (0.074)\\
 & 40 & 0.001 (0.003) & 0.004 (0.003) & 0.004 (0.003) & 0.145 (0.002) & 0.137 (0.002) & 0.150 (0.002) & 100.000 (0.000) & 100.000 (0.000) & 100.000 (0.000)\\
\bottomrule
\\\\
\multicolumn{11}{c}{$\bm{T_{pre}=15, T_{post}=5}$} \\
\toprule
& & \multicolumn{3}{c}{Bias} & \multicolumn{3}{c}{Empirical SE} & \multicolumn{3}{c}{Coverage (\%)} \\
\cmidrule(lr){3-5}\cmidrule(lr){6-8}\cmidrule(lr){9-11}
$N_{tr}$ & $N_{co}$ & ASCM & Part. pooled SCM & Part. pooled SCM & ASCM & Part. pooled SCM & Part. pooled SCM & ASCM & Part. pooled SCM & Part. pooled SCM \\
& & & $(\nu=\hat{\nu})$ & $(\nu=1)$ & & $(\nu=\hat{\nu})$ & $(\nu=1)$ & & $(\nu=\hat{\nu})$ & $(\nu=1)$ \\
\midrule
\multirow{4}{*}{1} & 6 & 0.001 (0.005) & -0.003 (0.004) & -0.003 (0.004) & 0.219 (0.004) & 0.190 (0.003) & 0.190 (0.003) & 95.211 (0.490) & 100.000 (0.000) & 100.000 (0.000)\\
 & 10 & -0.013 (0.005) & -0.010 (0.004) & -0.010 (0.004) & 0.197 (0.003) & 0.180 (0.003) & 0.180 (0.003) & 94.737 (0.512) & 100.000 (0.000) & 100.000 (0.000)\\
 & 25 & -0.002 (0.004) & 0.000 (0.004) & 0.000 (0.004) & 0.182 (0.003) & 0.174 (0.003) & 0.174 (0.003) & 95.000 (0.500) & 100.000 (0.000) & 100.000 (0.000)\\
 & 40 & 0.003 (0.004) & 0.003 (0.004) & 0.003 (0.004) & 0.181 (0.003) & 0.177 (0.003) & 0.177 (0.003) & 93.579 (0.562) & 100.000 (0.000) & 100.000 (0.000)\\
\cmidrule{1-11}
\multirow{2}{*}{10} & 25 & -0.001 (0.002) & -0.001 (0.002) & -0.001 (0.002) & 0.079 (0.001) & 0.079 (0.001) & 0.071 (0.001) & 94.895 (0.505) & 99.579 (0.149) & 99.421 (0.174)\\
 & 40 & 0.004 (0.002) & 0.002 (0.002) & 0.002 (0.002) & 0.076 (0.001) & 0.076 (0.001) & 0.067 (0.001) & 95.526 (0.474) & 100.000 (0.000) & 100.000 (0.000)\\
\bottomrule
\\\\
\multicolumn{11}{c}{$\bm{T_{pre}=30, T_{post}=10}$} \\
\toprule
& & \multicolumn{3}{c}{Bias} & \multicolumn{3}{c}{Empirical SE} & \multicolumn{3}{c}{Coverage (\%)} \\
\cmidrule(lr){3-5}\cmidrule(lr){6-8}\cmidrule(lr){9-11}
$N_{tr}$ & $N_{co}$ & ASCM & Part. pooled SCM & Part. pooled SCM & ASCM & Part. pooled SCM & Part. pooled SCM & ASCM & Part. pooled SCM & Part. pooled SCM \\
& & & $(\nu=\hat{\nu})$ & $(\nu=1)$ & & $(\nu=\hat{\nu})$ & $(\nu=1)$ & & $(\nu=\hat{\nu})$ & $(\nu=1)$ \\
\midrule
\multirow{4}{*}{1} & 6 & 0.000 (0.004) & -0.002 (0.003) & -0.002 (0.003) & 0.155 (0.003) & 0.125 (0.002) & 0.125 (0.002) & 95.579 (0.472) & 100.000 (0.000) & 100.000 (0.000)\\
 & 10 & -0.000 (0.003) & -0.001 (0.003) & -0.001 (0.003) & 0.146 (0.002) & 0.128 (0.002) & 0.128 (0.002) & 94.474 (0.524) & 100.000 (0.000) & 100.000 (0.000)\\
 & 25 & 0.007 (0.003) & 0.003 (0.003) & 0.003 (0.003) & 0.129 (0.002) & 0.121 (0.002) & 0.121 (0.002) & 95.684 (0.466) & 100.000 (0.000) & 100.000 (0.000)\\
 & 40 & -0.001 (0.003) & -0.002 (0.003) & -0.002 (0.003) & 0.129 (0.002) & 0.122 (0.002) & 0.122 (0.002) & 94.000 (0.545) & 100.000 (0.000) & 100.000 (0.000)\\
\cmidrule{1-11}
\multirow{2}{*}{10} & 25 & 0.001 (0.001) & 0.000 (0.001) & 0.000 (0.001) & 0.051 (0.001) & 0.050 (0.001) & 0.047 (0.001) & 93.526 (0.565) & 99.789 (0.105) & 99.895 (0.074)\\
 & 40 & 0.001 (0.001) & -0.001 (0.001) & -0.000 (0.001) & 0.051 (0.001) & 0.050 (0.001) & 0.046 (0.001) & 95.158 (0.492) & 100.000 (0.000) & 100.000 (0.000)\\
\bottomrule
\end{tabular}
\begin{tablenotes}
   \item The bias and empirical standard error of each estimator is divided by the standard deviation of the outcome variable after demeaning the outcome within each year. 
\end{tablenotes}
\end{threeparttable}
}
\end{table}
\end{landscape}

\end{document}